\documentclass[lettersize,journal]{IEEEtran}

\usepackage{amsmath,amsfonts,amssymb}

\usepackage{algorithmic}
\usepackage{algorithm}
\usepackage{array}
\usepackage{graphicx}
\usepackage{subcaption}
\usepackage[caption=false,font=normalsize,labelfont=sf,textfont=sf]{subfig}
\usepackage{textcomp}
\usepackage{stfloats}
\usepackage{url}

\usepackage{verbatim}
\usepackage{graphicx}
\usepackage{xcolor}
\usepackage{cleveref}
\usepackage{multirow}
\usepackage{cite}
\usepackage{cuted}
\usepackage{lipsum} 
    \allowdisplaybreaks
\begin{document}

\title{Performance Optimization of a RIS-Assisted PLC–RF–UOWC System with Imperfect CSI
}

\author{Md. Ibrahim, and Md. Abdur Rakib

\thanks{

M. Ibrahim is with the Institute of ICT, Rajshahi University of Engineering \& Technology (RUET) (email: ibrahimruet.007@gmail.com)

M. Abdur Rakib is with the Department of EEE,  RUET (email: arafatrakib1310@gmail.com) (\textit{corresponding author: Md. Abdur Rakib})
}}

\maketitle 

\begin{abstract}
This paper proposes a novel reconfigurable intelligent surface (RIS)-assisted three-hop power line communication (PLC)–radio frequency (RF)–underwater optical wireless communication (UOWC) hybrid system for reliable indoor-to-underwater connectivity. Considering imperfect channel state information, closed-form expressions for the outage probability and average bit error rate are derived. The study investigates RIS element and placement optimization, revealing that the UOWC link requires more reflecting elements than the RF link to compensate for severe underwater turbulence. Furthermore, the results show that placing the RIS at the midpoint between the source and destination requires a larger number of reflecting elements, whereas optimized placement enhances beamforming gain and improves overall system reliability with fewer elements.
\end{abstract}


\section{Introduction}

The rapid evolution of sixth-generation (6G) wireless networks requires integrated communication architectures capable of overcoming the limitations of conventional radio frequency (RF) systems \cite{padhan2025advancing}, particularly in terrestrial and underwater domains. In this context, power line communication (PLC) offers a cost-effective backbone for smart grid connectivity \cite{yang2021performance}, while reconfigurable intelligent surfaces (RISs) can significantly enhance RF and underwater optical wireless communication (UOWC) links by dynamically controlling electromagnetic and optical wave propagation to mitigate scattering and turbulence \cite{9684924,10210651}. Recent studies have investigated various dual-hop communication architectures, including PLC–RF systems under impulsive noise and hardware impairments \cite{yang2021performance}, hybrid FSO–UOWC frameworks considering atmospheric and oceanic effects \cite{10535219}, VLC–RF–PLC indoor systems for energy efficiency \cite{8862946}, and mixed RF–UAC in \cite{10499864} or PLC–FSO relaying networks in \cite{9750196}. However, most of these works either neglect RIS assistance, imperfect channel state information (CSI), or focus on limited propagation environments.

Several studies have investigated RIS-assisted communication systems under different configurations and propagation environments. For example, \cite{9684924} developed an RIS-assisted model considering environmental effects, while \cite{10210651} proposed an UOWC system showing performance improvements. Performance analysis of RIS-assisted RF–UOWC and THz–UOWC systems was presented in \cite{9352962,rakib2024ris}; however, PLC integration was not considered. A PLC–RF system with closed-form outage probability (OP) and average bit error rate (ABER) under impulsive noise was studied in \cite{10086658} without RIS assistance. Moreover, \cite{10025392} and \cite{10702588} investigated imperfect CSI in RIS-aided networks but did not consider PLC–UOWC architectures. Finally, a RIS-enhanced PLC–RF system was analyzed in \cite{9513286}, although the study was limited to terrestrial RF links and did not include imperfect CSI modeling.


Although several studies have investigated hybrid dual-hop communication systems \cite{10535219,10499864}, they primarily focus on two-hop architectures and do not incorporate PLC links for indoor connectivity. Furthermore, many existing works either neglect RIS deployment \cite{10499864,9750196} or assume ideal channel conditions with perfect CSI \cite{9684924,10210651,9352962,rakib2024ris}, which limits their applicability to practical environments. To overcome these limitations, this paper proposes a novel hybrid communication architecture for indoor-to-underwater connectivity in coastal and offshore monitoring applications. The proposed framework employs RIS assistance in both RF and UOWC links to mitigate channel impairments assuming imperfect CSI. Furthermore, the optimal number of RIS elements and their optimal placement are jointly investigated to enhance system reliability. To the best of the authors' knowledge, this work presents the first performance analysis of an RIS-assisted PLC–RF–UOWC communication system that investigates both RIS element optimization and placement optimization.


\section{System Model and Problem Formulation}
We consider a hybrid communication architecture for coastal and offshore monitoring, where reliable transmission is required between an indoor control unit and an underwater node. The source node $\mathcal{S}$ is located inside a building and transmits data to a rooftop gateway $\mathcal{G}$ through a PLC link using the existing electrical wiring network. The gateway then forwards the signal to a relay $\mathcal{R}$ (e.g., a coastal tower or buoy station). Due to blockage from surrounding structures, the direct RF link between $\mathcal{G}$ and $\mathcal{R}$ is unavailable. Therefore, a RIS deployed on a nearby structure establishes an indirect reflected RF link. The relay converts the received RF signal into an optical signal and forwards it to the underwater destination $\mathcal{D}$. Since direct optical transmission may be degraded by underwater obstacles, an underwater RIS is deployed to assist the UOWC link.
We consider that the PLC link is subject to log-normal fading with mean $\mu_p$ and variance $\sigma_p^{2}$. The instantaneous SNR is expressed as $\gamma_p = \frac{E_p |h_p|^2}{N_p}$, where $E_p$ represents the average transmitted signal energy and $N_p$ denotes the noise power. Specifically, $N_p = \sigma_p^{2}$ for background noise (BN), whereas $N_p = \sigma_p^{2}(1+\eta_{p})$ in the presence of impulsive noise (IN), with $\eta$ representing the ratio of IN power to BN power. Therefore, the CDF of $\gamma_p$ is expressed as \cite[Eqs.~(7)]{yang2021performance}
\begin{align}
  \label{cdf plc}
    F_{\gamma_{p}}(\gamma)=&\frac{1-P_n}{\Gamma(m_{p})}G_{1,2}^{1,1}\biggl[\frac{m_{p}\gamma}{\Omega_p}\bigg| 
\begin{array}{c}
 1 \\
m_{p},0 \\
\end{array}
\biggl]
\nonumber
\\
+&\frac{P_n}{\Gamma(m'_{p})}G_{1,2}^{1,1}\biggl[\frac{m'_{p}\gamma}{\Omega'_p}\bigg| 
\begin{array}{c}
 1 \\
 m'_{p},0 \\
\end{array}
\biggl], 
\end{align}
where
$P_n$ denotes the probability of IN occurrence, $m_{p}=m'_{p}=\frac{1}{e^{4\sigma^2_p}-1}$ denote the shadowing severity, $\Omega_p=\Bar{\gamma}_{p}e^{2\mu_p}\sqrt{(1+m_{p})/m_{p}}$ and $\Omega'_p=\Bar{\gamma}'_{p}e^{2\mu_p}\sqrt{(1+m'_{p})/m'_p}$ are the average power, $\Bar{\gamma}_{p}=\frac{E_p}{\sigma_p^2}$, and $\Bar{\gamma}'_{p}=\frac{E_p}{\sigma_p^2(1+\eta_{p})}$.

Since the direct link between $\mathcal{G}$ and $\mathcal{R}$ is blocked, a RIS denoted by $\mathcal{T}_r$ is deployed to enable communication. The RF channels of the $\mathcal{G}$–$\mathcal{T}_r$ and $\mathcal{T}_r$–$\mathcal{R}$ links follow independent $\kappa$–$\mu$ fading distributions. Let $\alpha_i$ and $\beta_i$ represent the channel amplitudes of the $\mathcal{G}$–$\mathcal{T}_r$ and $\mathcal{T}_r$–$\mathcal{R}$ links corresponding to the $i$th RIS element. Perfect CSI is assumed for the $\mathcal{G}$–$\mathcal{T}_r$ link, while the $\mathcal{T}_r$–$\mathcal{R}$ link experiences imperfect CSI modeled as $\beta_i=\rho\widehat{\beta}_i+\sqrt{1-\rho^2}\Delta\beta_i$, where $\rho$ is the correlation coefficient and $\Delta\beta_i$ denotes the estimation error. Under these assumptions, the instantaneous SNR at the relay is given by
$\gamma_r=|\sum^{N_{r}}_{i=1}\alpha_i\widehat{\beta_i}e^{(\phi_i-\theta_i-\psi_i)}|^2\rho^2\Bar{\gamma}_i$, where $N_r$ is the number of RIS elements and $\bar{\gamma}_r$ is the average RF SNR. By aligning the reflected signals through $\phi_i=\theta_i+\psi_i$, the received signals are coherently combined, resulting in the maximum instantaneous SNR at $\mathcal{R}$ and is expressed as $\gamma_r=(\sum^{N_{r}}_{i=1}|\alpha_i||\widehat{\beta_i}|)^2\rho^2\bar{\gamma}_r$.
Therefore, the CDF expression of $\gamma_r$ can be expressed as 
\begin{align}
\label{cdf rf}
   F_{\gamma_r}(\gamma)=&\frac{\chi_1\gamma^{w_{r}}}{\Gamma(b_{r})\Gamma(w_{r})\nu_{r}^{w_{r}}}G_{2,2}^{1,2}\left[\chi_2\gamma\Bigg| 
\begin{array}{c}
 \chi_3,1-w_{r} \\
 0,-w_{r} \\
\end{array}
\right],
\end{align}
where $\chi_1=(\frac{L_r}{\rho^{2}a_{r}})^{w_{r}}$,
$\chi_2=\frac{L_r}{a_{r}\rho^{2}{\nu_{r}}}$,
$\chi_3=1-b_{r}-w_{r}$,
$a_{r}=\frac{1}{\sigma_{r}}(1+\frac{1}{\sigma_{r}\Bar{\gamma}_r})$,
$b_{r}=(1+\frac{1}{\sigma_{r}\Bar{\gamma}_r})^2$, \textcolor{black}{$L_r=\Biggl\{\biggl(\frac{\lambda_r}{4\pi}\biggl)^{4}\frac{G_{r,1}G_{r,2}\zeta_{r}}{d_{r,1}^{2}d_{r,2}^{2}}\Biggl\}^{-1}$ is the path loss, $\lambda_r$ denotes the wavelength, $G_{r,1}$ and $ G_{r,2}$ represent the RIS channel gains for $\mathcal{G}$–$\mathcal{T}_r$ and $\mathcal{T}_r$–$\mathcal{R}$ links, respectively, $d_{r,1}$ and $d_{r,2}$ correspond to the distances from $\mathcal{G}$ to $\mathcal{T}_r$ and from $\mathcal{T}_r$ to $\mathcal{R}$, respectively, $\zeta_r$ is the RIS efficiency}, $w_{r}$ and $\nu_{r}$ denote the shape and scale parameters, respectively.
\textbf{Proof}: See Appendix.

Similar to the RF hop, the direct link between $\mathcal{R}$ and $\mathcal{D}$ is unavailable due to severe attenuation in underwater environments. Therefore, a RIS denoted by $\mathcal{T}_u$ is deployed to establish an indirect UOWC link. The $\mathcal{R}$–$\mathcal{T}_u$ and $\mathcal{T}_u$–$\mathcal{D}$ channels are assumed to follow the mEGG distribution. Let $\alpha_n$ and $\beta_n$ denote the channel amplitudes of the $\mathcal{R}$–$\mathcal{T}_u$ and $\mathcal{T}_u$–$\mathcal{D}$ links corresponding to the $n$th RIS element. The instantaneous SNR at $\mathcal{D}$ is expressed as
$\gamma_u=|\sum^{N_{u}}_{n=1}\alpha_n\beta_ne^{(\phi_n-\theta_n-\psi_n)}|^2\Bar{\gamma}_u$, where $N_u$ is the number of RIS elements and $\bar{\gamma}_u$ denotes the average UOWC SNR. By setting $\phi_n=\theta_n+\psi_n$, coherent phase alignment is achieved, yielding $\gamma_r=(\sum^{N_{u}}_{n=1}|\alpha_n||\beta_n|)^2\Bar{\gamma}_u$.
Based on this expression, the CDF expression of $\gamma_u$ can be obtained as \cite[Eqs. (22)]{rakib2024ris} 
\begin{align}
\label{cdf uw}
F_{\gamma_{u}}
(\gamma)=&\frac{\mathcal{B}_{1}\gamma^{\frac{w_{u}}{s}}}{(2\pi)^{\frac{s-1}{2}}}
G_{1,s+1}^{s,1}\left[\frac{L_{u}\mathbb{E}(V_{i})^{s}\gamma}{\Bar{\gamma}_o(w_{u}s)^{s}}\biggl | 
    \begin{array}{c}
     1-\frac{w_{u}}{s} \\
     0,\frac{s-1}{s},-\frac{w_{u}}{s} \\ 
    \end{array}
    \right],
\end{align}

where $\mathcal{B}_{1}=\frac{\mathbb{E}(V_{i})^{w_{u}}L_{u}^{\frac{w_{u}}{s}}}{\sqrt{s}\Gamma(w_{u})\nu_{u}^{w_{u}}\Bar{\gamma}_o^{\frac{w_{u}}{s}}}$, $w_u=\frac{N_{u}\mathbb{E}(V_{i})^{2}}{Var(V_{i})}$ is the shape parameter, $\nu_u=\frac{Var(V_{i})}{{\mathbb{E}(V_{i})}}$ is the scale parameter, $s$ represents the detection signal method ($s=1$ corresponds to HD detection and $s=2$ indicates the IM/DD detection), $\Bar{\gamma}_{o}=\frac{(\varrho_{0}\mathbb{E}(V_{i}))^{s}}{\varepsilon}$ denotes the electrical SNR, $\varrho$ is the conversion co-efficient, $\varepsilon$ is the variance of noise, \textcolor{black}{$L_{u}=\Biggl\{\biggl(\frac{\lambda_o}{4\pi}\biggl)^{4}\frac{G_{u,1}G_{u,2}\zeta_{u}}{d_{u,1}^{2}d_{u,2}^{2}}\Biggl\}^{-1}$ denotes the path loss, $\lambda_u$, $G_{u,1}$, $ G_{u,2}$, $d_{u,1}$, $d_{u,2}$, and $\zeta_u$ have the same definitions as described earlier but for the $\mathcal{R}$ to $\mathcal{T}_u$ and from $\mathcal{T}_u$ to $\mathcal{D}$, respectively}. Now, the $k^{th}$ moment of random variable $V_i$ is obtained as $  \mathbb{E}(V_{i}^{k})=\frac{\mathcal{I}_1 \mathcal{I}_{2}^{k}\prod_{i=1}^{4}\Gamma(k+\mathcal{M}_{1,i})}{\prod_{i=1}^{2}\Gamma(k+\mathcal{C}_{1,i})}
   +\frac{\mathcal{I}_3(\mathcal{I}_4)^{kc'_{u}}\prod_{i=1}^{4}\Gamma\left(k+\mathcal{M}_{2,i}\right)}{\prod_{i=1}^{2}\Gamma\left(k+\mathcal{C}_{2,i}\right)}+
\frac{\mathcal{I}_5(\mathcal{I}_6)^{kc_{u}}\prod_{i=1}^{4}\Gamma(k+\mathcal{M}_{3,i})}{\prod_{i=1}^{2}\Gamma(k+\mathcal{C}_{3,i})}+\frac{\mathcal{I}_{7}(\iota_{u}b_{u})^{kc_{u}}(\iota'_{u}b'_{u})^{kc'_{u}}\prod_{i=1}^{4}\Gamma\left(k+\mathcal{M}_{4,i}\right)}{\prod_{i=1}^{2}\Gamma\left(k+\mathcal{C}_{4,i}\right)}$, where $\mathcal{I}_1=\omega_{u}\omega'_{u}\xi_{u}^{2}{\xi}_{u}'^{2}$, 
$\mathcal{I}_2=\iota_{u}\iota'_{u}\lambda_{u}\lambda'_{u}$, 
$\mathcal{I}_3=\frac{\omega_{u}(1-\omega'_{u})\xi_{u}^{2}{\xi_{u}'}^{2}}{(2\pi)^{\frac{1}{2}(c'_{u}-1)}{c_{u}'}^{\frac{1}{2}}\Gamma(a'_{u})}$,
$\mathcal{I}_4=\iota_{u}\iota'_{u}b'_{u}c'_{u}\lambda_{u}$,
$\mathcal{I}_5=\frac{\omega'_{u}(1-\omega_{u})\xi_{u}^{2}{\xi_{u}'}^{2}}{(2\pi)^{\frac{1}{2}(c_{u}-1)}c_{u}^{\frac{1}{2}}\Gamma(a_{u})}$,
$\mathcal{I}_6=\iota'_{u}\iota_{u}b_{u}c_{u}\lambda'_{u}$,
$\mathcal{I}_7=\frac{(1-\omega_{u})(1-\omega'_{u})\xi_{u}^{2}{\xi_{u}'}^{2}}{\Gamma(a_{u})\Gamma(a'_{u})}$, $\mathcal{M}_{1}=\begin{bmatrix}
    1 & {\xi'_{u}}^{2} & 1 & \xi_{u}^{2}
\end{bmatrix}$, $\mathcal{M}_{2}=\begin{bmatrix}
    a'_{u} & \frac{{\xi_{u}'}^{2}}{c'_{u}} & 1-\mathcal{A}_{1} & 1-\mathcal{A}_{2}
\end{bmatrix}$, $\mathcal{M}_{3}=\begin{bmatrix}
    a_{u} & \frac{\xi_{u}^{2}}{c_{u}} & \mathcal{A}_{3} & \mathcal{A}_{4}
\end{bmatrix}$, $\mathcal{M}_{4}=\begin{bmatrix}
    a'_{u} & \frac{{\xi_{u}'}^{2}}{c'_{u}} & a_{u} & \frac{\xi_{u}^{2}}{c_{u}}
\end{bmatrix}$, $\mathcal{C}_{1}=\begin{bmatrix}
    {\xi_{u}'}^{2}+1 & \xi_{u}^{2}+1
\end{bmatrix}$, $\mathcal{C}_{2}=\begin{bmatrix}
    \frac{{\xi_{u}'}^{2}}{c'_{u}}+1 & 1-\mathcal{A}_{5}
\end{bmatrix}$, $\mathcal{C}_{3}=\begin{bmatrix}
    \frac{\xi_{u}^{2}}{c_{u}}+1 & \mathcal{A}_{6}
\end{bmatrix}$, and $\mathcal{C}_{4}=\begin{bmatrix}
    \frac{{\xi_{u}'}^{2}}{c'_{u}}+1 & \frac{\xi_{u}^{2}}{c_{u}}+1
\end{bmatrix}$, $\mathcal{A}_{1}=0,...,\frac{c'_{u}-1}{c'_{u}}$, 
$\mathcal{A}_{2}=\frac{1-\xi_{u}^{2}}{c'_{u}},...,\frac{c'_{u}-\xi_{u}^{2}}{c'_{u}}$, 
$\mathcal{A}_{3}=\frac{1}{c_{u}},...,1$,
$\mathcal{A}_{4}=\frac{{\xi_{u}'}^{2}}{c_{u}},...,\frac{{\xi_{u}'}^{2}+c_{u}-1}{c_{u}}$,
$\mathcal{A}_{5}=-\frac{\xi_{u}^{2}}{c'_{u}},...,\frac{c'_{u}-\xi_{u}^{2}-1}{c'_{u}}$,
$\mathcal{A}_{6}=\frac{1+{\xi_{u}'}^{2}}{c_{u}},...,\frac{c_{u}+{\xi_{u}'}^{2}}{c_{u}}$, $a$, $b$, and $c$ denote the parameters of the GG distribution, $\omega$ is the mixture weight parameter, $\lambda$ corresponds to the exponential distribution parameter, and $\xi$ represents the pointing error of the respective links.

\section{Performance Analysis}
\subsection{OP Analysis}
Utilizing the decode-and-forward (DF) relay at $\mathcal{R}$, the instantaneous SNR can be expressed as $\gamma_{eq}=\min(\gamma_{p},\gamma_{r},\gamma_{u})$. Accordingly, the CDF of $\gamma_{eq}$ can be obtained as  
\begin{align}
    \label{cdfeq}
    F_{\gamma_{eq}}(\gamma)=1-\bigl\{(1- F_{\gamma_{p}}(\gamma))(1- F_{\gamma_{r}}(\gamma))(1- F_{\gamma_{u}}(\gamma))\bigl\}.
\end{align}
Now, substituting \eqref{cdf plc}, \eqref{cdf rf}, and \eqref{cdf uw} into \eqref{cdfeq}, the CDF of $\gamma_{eq}$ is obtained. Please note that a system outage occurs when the instantaneous SNR, $\gamma_{eq}$, falls below a predefined threshold SNR, $\gamma_{th}$. Therefore, OP can be mathematically expressed as $P_{out}=P_r \left\{\gamma_{eq} \leq \gamma _{\text{th}}\right\}=F_{\gamma_{eq}} (\gamma _{\text{th}}).$
By substituting \eqref{cdfeq}, the expression of OP is derived. Now, applying the identity of \cite[Eq.~(8.2.2.14)]{prudnikov1986integrals} and performing some mathematical manipulations, the asymptotic OP for the PLC link can be written as
\begin{align}
    \label{asy1}
    P_{out}^{plc}(\infty)=&\frac{1-P_n}{\Gamma(m_{p})}\frac{\Gamma(m_{p})}{\Gamma(1+m_{p})}\Bigl(\frac{\Omega_p}{m_{p}\gamma}\Bigl)^{-m_{p}}
\nonumber
\\
+&\frac{P_n}{\Gamma(m'_{p})}\frac{\Gamma(m'_{p})}{\Gamma(1+m'_{p})}\Bigl(\frac{\Omega'_p}{m'_{p}\gamma}\Bigl)^{-m'_{p}}.
\end{align}
Similarly, using the same identity, the asymptotic OP for the RF link and UOWC can be derived, respectively. In the proposed model, when the PLC link dominates due to severe shadowing, the first hop limits the system performance and the asymptotic OP is expressed as $P_{out}^{asy}(\infty)=(G_c^{p}\bar{\gamma})^{-G_d^{p}}$, where $G_c^{p}$ denotes the coding gain and $G_d^p=\min\{m_{p},m'_{p}\}$ denotes the diversity order. When the RF link becomes dominant, the second hop determines the system behavior with diversity order $G_d^{r}=w_r$. Finally, when the UOWC link dominates, the third hop governs the overall performance with diversity order $G_d^w=\min\left\{\frac{w_u}{s},\frac{s-1}{s}\right\}$.
\subsection{ABER Analysis}
The ABER of the proposed system can be expressed as
\begin{align}
    \label{aber}
    P_e=1-\bigl\{(1-p_{e_{1}})(1-p_{e_{2}})(1-p_{e_{3}})\bigl\},
\end{align}
where $p_{e_{1}}$, $p_{e_{2}}$, and $p_{e_{3}}$ denote the ABER of the PLC, RF, and UOWC links, respectively. However, the mathematical expression of ABER can be expressed as \cite[Eq. (41)]{rakib2024ris} 
\begin{align}
    \label{gaber}
    P_e=\frac{q^{p}}{2\Gamma(p)}\int_0^\infty e^{-q\gamma}\gamma^{p-1}F_{\gamma}(\gamma)d\gamma.
\end{align}
Substituting \eqref{cdf plc} into \eqref{gaber} and using the identity \cite[Eqs. (7.813.1)]{ edition2007table}, $P_{e_1}$ is derived as
\begin{align}
    p_{e_1}=&\frac{1-P_n}{2\Gamma(p)\Gamma(m_{p})}G_{2,2}^{1,2}\left[\frac{m_{p}}{\Omega_p q}\Bigg| 
\begin{array}{c}
1-p, 1 \\
m_{p},0 \\
\end{array}
\right]
\nonumber
\\ \nonumber
+& \frac{P_n}{2\Gamma(p)\Gamma(m'_{p})}G_{2,2}^{1,2}\left[\frac{m'_{p}}{\Omega'_{p}q}\Bigg| 
\begin{array}{c}
1-p, 1 \\
 m'_{p},0 \\
\end{array}
\right]
\end{align}
Now, similar to $p_{e_1}$, $p_{e_2}$ is derived as 
\begin{align} \nonumber
    p_{e_2}=&\frac{\mathcal{L}_{1}(q)^{-w_{r}}}{2\Gamma(p)}G_{3,2}^{1,3}\left[\frac{\chi_2}{q}\Bigg| 
\begin{array}{c}
 \chi_{10},\chi_3,1-w_{r} \\
 0,-w_{r} \\
\end{array}
\right],
\end{align}
where $\mathcal{L}_{1}=\frac{\chi_1}{\Gamma(b_{r})\Gamma(w_{r})\nu_{r}^{w_{r}}}$ and $\chi_{10}=1-p-w_c$. Finally, $p_{e_3}$ is obtained as
\begin{align} \nonumber
p_{e_3}=\frac{\mathcal{L}_2}{2\Gamma(p)q^{\frac{w_u}{s}}}
G_{2,s+1}^{s,2}\left[\frac{L_{u}\mathbb{E}(V_{i})^{s}}{q\Bar{\gamma}_o(w_{u}s)^{s}}\biggl | 
    \begin{array}{c}
    \mathcal{B}_{2}, 1-\frac{w_{u}}{s} \\
     0,\frac{s-1}{s},-\frac{w_{u}}{s} \\ 
    \end{array}
    \right],
\end{align}
where $\mathcal{L}_2=\frac{\mathcal{B}_{1}}{(\sqrt{2\pi})^{s-1}}$ and $\mathcal{B}_{2}=1-p-\frac{w_u}{s}$. Furthermore, following a similar approach to that used in deriving the asymptotic OP, the asymptotic ABER expressions for the PLC, RF, and UOWC links can also be derived.

\subsection{Optimization of RIS Reflecting Elements}
This subsection finds the minimum RIS
elements at the RF-hop and UOWC-hop RISs that satisfy
an OP target $P^{th}_{out}$. 

\subsubsection{Upper bound on the asymptotic OP}
Adopting DF relaying, the end-to-end OP satisfies $P_{out}=1-\big(1-P^{plc}_{out}\big)\big(1-P^{rf}_{out}\big)\big(1-P^{uw}_{out}\big)\le P^{plc}_{out}+P^{rf}_{out}+P^{uw}_{out}$. The PLC term does not depend on $N_r$ or $N_u$ and is hence considered as
a constant. The asymptotic expressions of RF and UOWC link contain Gamma function, which are difficult to handle in an
optimization problem. To obtain tractable expressions, we derive upper-bound asymptotic
OP expressions of the RF and UOWC links as $\tilde P^{rf}_{out}(\infty)=\chi_1\Big(\frac{e\gamma}{N_r\delta\nu_r}\Big)^{N_r\delta}\Big(b_r+N_r\delta\Big)^{N_r\delta}$ and $\tilde P^{uw}_{out}(\infty)=\Big(\frac{e\,\Phi}{N_u\lambda}\Big)^{N_u\lambda}$. Where $\delta=E(Y_i)^2/\mathrm{Var}(Y_i)$, $\lambda=E(V_i)^2/\mathrm{Var}(V_i)$, and
$\Phi=\frac{E(V_i)}{\nu_u}\big(\frac{L_u\gamma}{\bar\gamma_o}\big)^{1/s}$.

\subsubsection{Problem formulation}
In terms of reflecting elements, the end-to-end upper-bound asymptotic OP expression can
be written as $ \tilde P_{out}(N_r,N_u)=a+e^{f_r(N_r)}+e^{f_u(N_u)}$, where $a=P^{plc}_{out}(\infty)$. The terms  $f_r(N_r)\triangleq\ln\tilde P^{rf}_{out}(\infty)$
and $f_u(N_u)\triangleq\ln\tilde P^{uw}_{out}(\infty)$ are obtained as $f_r(N_r)=\ln \chi_1+A\delta N_r-\delta N_r\ln N_r+\delta N_r\ln\Big(b_r+\delta N_r\Big)$ and $f_u(N_u)=\lambda N_u+\lambda N_u\ln\Phi-\lambda N_u\ln(\lambda N_u)$, where $A=1+\ln \gamma-\ln(\delta\nu_r)$. Therefore, the optimization
problem is formulated as
\begin{subequations}\label{eq:P1}
\begin{align}
\min_{\mathbf{N}}\;& N_r+N_u\\
\text{s.t.}\;& 1\le N_r,N_u\le N_{max},\label{eq:P1b}\\
& a+e^{f_r(N_r)}+e^{f_u(N_u)}\le P^{th}_{out},\label{eq:P1c}\\
& N_r,N_u\in\mathbb{Z}^+.\label{eq:P1d}
\end{align}
\end{subequations}
The integer constraint \eqref{eq:P1d} is relaxed and the solution is
recovered by applying the ceiling function to the optimized
$\mathbf{N}$. Although both  $f_r(N_r)$ and $f_u(N_u)$ are concave in $N_r$ and $N_u$, respectively ($f_r''(N_r)=-\frac{\delta b_r^2}{N_r(b_r+\delta N_r)^2}<0$ and $f_u''(N_u)=-\frac{\lambda}{N_u}<0$), constraint \eqref{eq:P1c} is not guaranteed to be convex. We therefore
solve \eqref{eq:P1} with a path-following algorithm.

\subsubsection{Path-following algorithm}
Let $\mathbf{N}^{(\kappa)}=\{N_r^{(\kappa)},N_u^{(\kappa)}\}$ be the feasible
point from iteration $\kappa-1$. Since
$f_r$ and $f_u$ are concave, they are upper bounded by their tangents at
$\mathbf{N}^{(\kappa)}$ as $f_r(N_r)\le\alpha_r+\beta_rN_r$ and 
$ f_u(N_u)\le\alpha_u+\beta_uN_u$. Where $\beta_r=f_r'\big(N_r^{(\kappa)}\big)$, $\alpha_r=\ln \chi_1+\frac{\delta b_rN_r^{(\kappa)}}{b_r+\delta N_r^{(\kappa)}}$,
$\beta_u=\lambda\ln\frac{\Phi}{\lambda N_u^{(\kappa)}}$, and 
$\alpha_u=\lambda N_u^{(\kappa)}$. 
The next feasible
point $\mathbf{N}^{(\kappa+1)}$ is therefore generated by solving the
convex problem
\begin{subequations}\label{eq:P2}
\begin{align}
\min_{\mathbf{N}}\;& N_r+N_u\\
\text{s.t.}\;& 1\le N_r,N_u\le N_{max},\\
& a+e^{\alpha_r+\beta_rN_r}+e^{\alpha_u+\beta_uN_u}\le P^{th}_{out}.
\end{align}
\end{subequations}
The overall procedure is summarized in Algorithm~1.

\begin{algorithm}[t]
\caption{Algorithm for solving optimization problem \eqref{eq:P1}}
\begin{algorithmic}[1]
\STATE \textbf{Initialization:} Initialize a feasible point $\mathbf{N}^{(0)}=\{N_{max},N_{max}\}$. Set $\kappa=0$.
\STATE \textbf{Repeat until convergence of $N_r$ and $N_u$:} Compute $\alpha_r,\beta_r,\alpha_u,\beta_u$. Generate $\mathbf{N}^{(\kappa+1)}$ by solving \eqref{eq:P2}. Reset $\kappa\leftarrow\kappa+1$.
\STATE \textbf{Output:} $N_r^*=\lceil N_r^{(\kappa)}\rceil$ and $N_u^*=\lceil N_u^{(\kappa)}\rceil$.
\end{algorithmic}
\end{algorithm}

\begin{figure*}[t]
\centering
\begin{subfigure}[t]{0.24\textwidth}
\centering
\includegraphics[width=\linewidth]{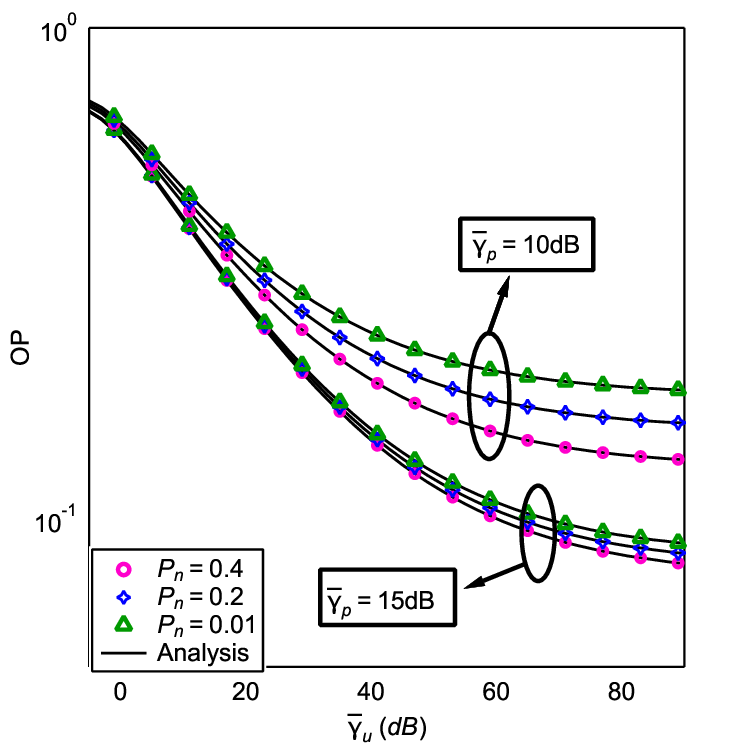}
\caption{OP versus $\bar{\gamma}_{u}$ for different values of $P_{n}$.}
\label{fig 1}
\end{subfigure}\hfill
\begin{subfigure}[t]{0.24\textwidth}
\centering
\includegraphics[width=\linewidth]{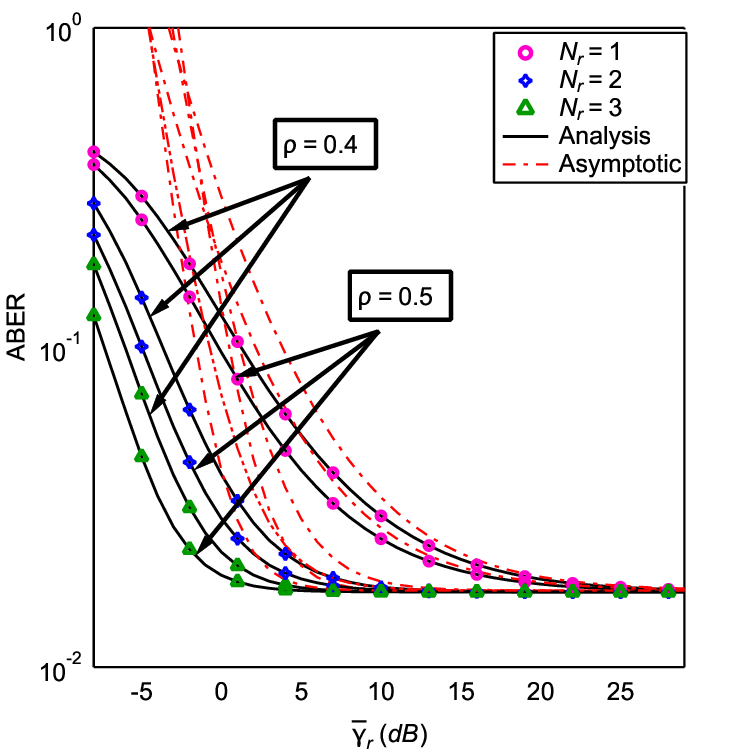}
\caption{ABER versus $\bar{\gamma}_{r}$ for different values of $N_{r}$ and $\rho$.}
\label{fig 3}
\end{subfigure}\hfill
\begin{subfigure}[t]{0.24\textwidth}
\centering
\includegraphics[width=\linewidth]{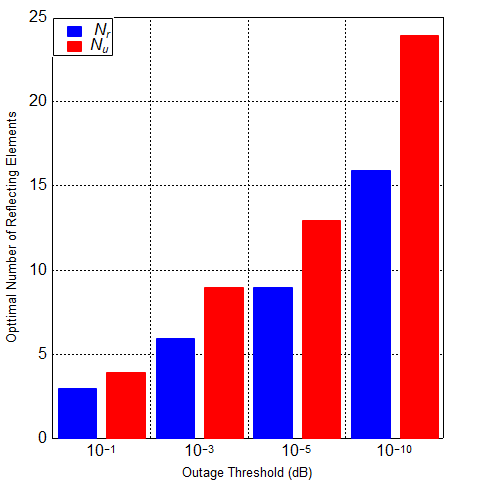}
\caption{Optimum number of reflecting elements versus outage threshold.}
\label{fig 4}
\end{subfigure}\hfill
\begin{subfigure}[t]{0.24\textwidth}
\centering
\includegraphics[width=\linewidth]{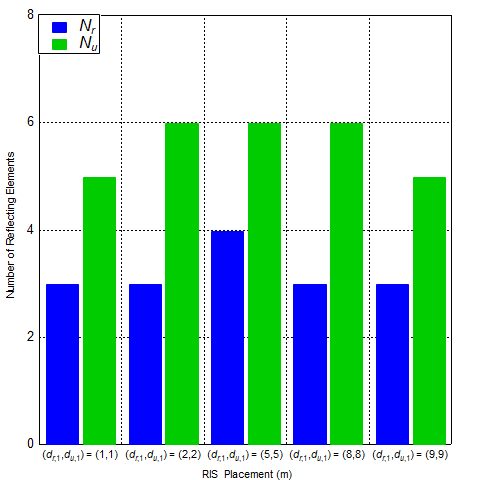}
\caption{Number of reflecting elements versus different RIS placement.}
\label{fig 2}
\end{subfigure}

\caption{Impacts of several system parameters on network performance.}
\label{fig:system_performance}
\end{figure*}
\subsection{Optimization of RIS Placement}
This subsection determines the positions of the RF-hop
RIS and the UOWC-hop RIS that minimize the
end-to-end OP.
\subsubsection{System geometry and decision variables}
The RF hop spans a distance $D_r$ between $\mathcal{G}$ and $\mathcal{R}$, and the UOWC hop
connects $\mathcal{R}$ and $\mathcal{D}$ over a distance $D_u$. Each RIS is positioned on the
straight line between its terminals, such that $d_{r,2}=D_r-d_{r,1}$ and
$d_{u,2}=D_u-d_{u,1}$. The optimization variables are
$\mathbf{d}\triangleq\{d_{r,1},d_{u,1}\}$. Since the path-loss model of $L_r$ and $L_u$ is valid only in the far
field, each RIS must be at least $d_{\min}$ away from both terminals.

\subsubsection{Problem formulation}
The asymptotic upper-bound OP in terms of distance can be written as $\tilde P_{out}(d_{r,1},d_{u,1})=a+z_r\,q_r^{2w_r}+z_u\,q_u^{\frac{2w_u}{s}}$. Where $z_r=\Big(\frac{c_r}{\rho^{2}a_r}\Big)^{w_r}
\Big(\frac{e\gamma}{w_r\nu_r}\Big)^{w_r}\Big(b_r+w_r\Big)^{w_r}$, 
$z_u=\biggl\{\frac{e}{w_u}\,\frac{\mathbb E(V_i)}{\nu_u}\Big(\frac{c_u\gamma}{\bar\gamma_o}\Big)^{\frac{1}{s}}\biggl\}^{w_u}$, $c_r=\frac{(4\pi/\lambda_{r})^{4}}{G_{r,1}G_{r,2}\zeta_r}$, and $c_u=\frac{(4\pi/\lambda_{o})^{4}}{G_{u,1}G_{u,2}\zeta_u}$, $q_r= d_{r,1}(D_r-d_{r,1})$, and $q_u= d_{u,1}(D_u-d_{u,1})$. The optimal placement problem is formulated as
\begin{subequations}\label{eq:P3}
\begin{align}
\min_{\mathbf{d}}\;&\tilde P_{out}(d_{r,1},d_{u,1})\\
\text{s.t.}\;& d_{\min}\le d_{r,1}\le D_r-d_{\min},\label{eq:P3b}\\
& d_{\min}\le d_{u,1}\le D_u-d_{\min}.\label{eq:P3c}
\end{align}
\end{subequations}
We rewrite the asymptotic upper-bound OP as $\tilde P_{out}=a+e^{f_r(d_{r,1})}+e^{f_u(d_{u,1})}$. The exponent terms can be obtained as $f_r(d_{r,1})=\ln z_r+2w_r\big\{\ln d_{r,1}+\ln(D_r-d_{r,1})\big\}$ and $f_u(d_{u,1})=\ln z_u+\frac{2w_u}{s}\big\{\ln d_{u,1}+\ln(D_u-d_{u,1})\big\}$.

\subsubsection{Optimal solution}
Since the exponential terms are increasing, \eqref{eq:P3} reduces to minimizing $f_r(d_{r,1})$ and $f_u(d_{u,1})$ separately. However, both $f_r(d_{r,1})$ and $f_u(d_{u,1})$ are concave in $d_{r,1}$ and $d_{u,1}$, respectively as
$f_r''(d_{r,1})=-2w_r\Big\{\frac{1}{d_{r,1}^2}+\frac{1}{(D_r-d_{r,1})^2}\Big\}<0$ and
$f_u''(d_{u,1})=-\frac{2w_u}{s}\Big\{\frac{1}{d_{u,1}^2}+\frac{1}{(D_u-d_{u,1})^2}\Big\}<0$. A concave function attains
its minimum at an end point of its interval. Hence the optimal placement
is
\begin{equation}
d_{j,1}^*\in\{d_{\min},\,D_j-d_{\min}\},\qquad j\in\{r,u\}.
\label{eq:dopt}
\end{equation}
Each RIS should therefore be placed as close as possible to one of its
terminals. Both end points give the same OP, since both functions are also symmetric
about the midpoint of their interval. Conversely, $f_j'(d_{j,1})=0$ at $d_{j,1}=D_j/2$. Since $f_j$ is concave, this
stationary point is a maximum. Therefore, placing a RIS at the midpoint of its hop
gives the highest outage probability. Problem \eqref{eq:P3} can be further solved using a path-following algorithm by replacing $f_r(d_{r,1})$ and $f_u(d_{u,1})$ with their respective first-order Taylor approximations around the current feasible point $\mathbf{d}^{(\kappa)}$ at the $\kappa$-th iteration. The resulting convex approximations are then solved iteratively to obtain the next feasible point $d_{j,1}^{(\kappa+1)}$, following the procedure outlined in Algorithm~1.

\section{Numerical Results}
  This section presents system performance results based on the derived OP and ABER. Analytical results are validated via Monte Carlo simulations with $10^6$ independent channel realizations. Unless stated otherwise, the parameters are set as $N_r=N_u=1$, $P_n=0.2$, $m_p=m'_p=8$, $\eta_p=15$, $\sigma_p=0.23$, $\kappa_r=\kappa'_r=2$, $\mu_r=\mu'_r=2$, $m_r=m'_r=1$, $\rho=0.7$, $s=(1,2)$, $\xi_u=\xi'_u=2$, $\bar{\gamma}_p=25$ dB, $\bar{\gamma}_r=10$ dB, and $\bar{\gamma}_u=30$ dB. Asymptotic analysis is also provided, demonstrating close agreement with both analytical and simulation results.

Fig. \ref{fig 1} illustrates the OP versus $\bar{\gamma}_u$ for different values of $P_n$. As $\bar{\gamma}_u$ increases from $0$ to about $40$ dB, the OP significantly decreases in all cases, indicating that system performance is mainly governed by the UOWC link in the low-to-mid SNR region. Moreover, reducing $P_n$ from $0.4$ to $0.01$ improves the OP from $3.28 \times 10^{-1}$ to $2.86 \times 10^{-1}$, corresponding to an improvement of about $14.69\%$ at $21$ dB. It is also observed that a higher PLC SNR (e.g., $\bar{\gamma}_p=15$ dB) results in lower OP. However, as $\bar{\gamma}_u$ exceeds $60$ dB, the OP curves approach an outage floor, indicating that the overall performance becomes limited by the PLC hop rather than the optical link.


Fig. \ref{fig 3} shows the ABER performance versus $\bar{\gamma}_r$ for different values of $\rho$ and $N_r$. The results indicate that increasing the number of RIS elements significantly improves performance. For example, increasing $N_r$ from $1$ to $3$ reduces the ABER from $2.2 \times 10^{-2}$ to $1.7 \times 10^{-2}$ at $15$ dB, corresponding to a $29.41\%$ improvement. This gain is attributed to enhanced beamforming capability, which strengthens signal focusing at the relay and mitigates RF multipath fading. Moreover, the impact of channel estimation accuracy is observed through $\rho$, where a higher value (e.g., $\rho=0.5$) provides better ABER performance due to more accurate RIS phase adjustments and improved coherent signal combining.

Fig. \ref{fig 4} illustrates the optimal number of RIS reflecting elements for the RF and UOWC links under different OP thresholds. It is observed that the required number of RIS elements increases as the OP requirement becomes more stringent (from $10^{-1}$ to $10^{-10}$). In particular, tightening the outage threshold increases the optimal number of reflecting elements from $3$ to $16$ for the RF link and from $4$ to $24$ for the UOWC link. This behavior indicates that lower OP requires stronger signal enhancement through higher beamforming gain. Furthermore, the UOWC link consistently requires more RIS elements than the RF link, highlighting the need for stronger mitigation against severe underwater turbulence. attenuation and pointing error.

Fig.~\ref{fig 2} presents the optimal number of RIS reflecting elements required under different RIS placement configurations for the RF and UOWC links. The results indicate that, to satisfy the target outage threshold of $p_{\mathrm{out}}^{\mathrm{th}}=10^{-1} dB$, the largest numbers of reflecting elements ($N_r=4$ and $N_u=6$) are required when the RISs are positioned midway along their respective 10 m links, i.e., $d_{r,1}=d_{u,1}=5$~m. In contrast, the required numbers decrease to $N_r=3$ and $N_u=5$ when the RF-link RIS is placed closer to either the source or the relay ($d_{r,1}=1$ or $9$~m), while the UOWC-link RIS is positioned closer to either the relay or the destination ($d_{u,1}=1$ or $9$~m). These results demonstrate that positioning the RIS near the source or relay in the RF link and near the relay or destination in the UOWC link minimizes the required number of reflecting elements while satisfying the target outage threshold.

\section{Conclusions}
This paper presented a novel RIS-assisted three-hop PLC–RF–UOWC framework for reliable indoor-to-underwater communication. The key contributions include the derivation of closed-form and asymptotic expressions for OP and ABER under imperfect CSI, as well as the optimization of RIS reflecting elements and their placement in both RF and UOWC links. The numerical results validate the analytical derivations and demonstrate the significant performance gains achieved through optimized RIS deployment. These findings highlight the potential of the proposed framework for reliable indoor-to-underwater connectivity.

\section*{APPENDIX}
Denoting $Y=(\sum^{N_{r}}_{i=1}|\alpha_i||\widehat{\beta_i}|)^{2}$ and 
utilizing the PDF of \cite{10499864} and the formula of \cite[Eq.~(13)]{rakib2024ris}, the PDF of $Y$ can be expressed as 
\begin{align}
\label{prpdf}
    f_Y(y)=&\Lambda_1y^{w_{r}-1}{}G_{0,1}^{1,0}\biggl[\frac{L y}{\nu_{r}}\bigg| 
\begin{array}{c}
- \\
 0 \\
\end{array}
\biggl],
\end{align}
 where $\Lambda_1=\frac{L_{r}^{w_{r}}}{\Gamma(w_{r})\nu_{r}^{w_{r}}}$, $w_{r}=\frac{N_r\mathbb{E}(Y)^2}{Var(Y)}$ and $\nu_{r}=\frac{Var(Y)}{\mathbb{E}(Y)}$ are related to $k$th moment of random variable, $Y$ and is expressed as $\mathbb{E}(Y^{k})=\sum_{l_1=0}^{\infty}\sum_{l_2=0}^{\infty}\mathcal{L}_3\Gamma(\mu_s+l_1+k/2)\Gamma(\mu_r+l_2+k/2)$, where 
$\mathcal{L}_3=\frac{\Lambda_2\Lambda_3(\Lambda_4)^{\frac{\Lambda_5}{2}}}{{(\Lambda_6)^{(\Lambda_7+k)} }}$, $\Lambda_2=\frac{4\mu_{r}^{\mu_{r}}\mu_{r}'^{\mu_{r}'}m_{r}^{m_r}{m_{r}'}^{m'_r}(1+\kappa_{r})^{\mu_r}(1+\kappa'_r)^{\mu'_r}}{(r_r)^{2\mu_r}({r'_{r}})^{2\mu'_r}\Gamma(\mu_r)\Gamma(\mu'_r)(\mu_r\kappa_{r}+m_r)^{m_r}(\mu'_r\kappa'_r+m'_r)^{m'_r}}$,
$\Lambda_3=\frac{\Gamma(\mu_r)\Gamma(\mu'_r)\Gamma(m_r+l_1)\Gamma(m'_r+l_1)(\Lambda_8)^{l_1}(\Lambda_9)^{l_2}}{l_1!l_2!\Gamma(m_r)\Gamma(m'_r)\Gamma(\mu_r+l_1)\Gamma(\mu'_r+l_2)}$,
$\Lambda_4=\frac{\mu'_r(1+\kappa'_r)r_r^2}{\mu_r(1+\kappa_{r}){r'_{r}}^2}$,
$\Lambda_5=\mu_r-\mu'_r+l_1-l_2$,
$\Lambda_6=\frac{\sqrt{\mu_r\mu'_r(1+\kappa_{r})(1+\kappa'_r)}}{r_{r}{r'_{r}}}$,
$\Lambda_7=\mu_r+\mu'_r+l_1+l_2$, 
$\Lambda_8=\frac{\mu_r^2\kappa_{r}(1+\kappa_{r})}{(\mu_r\kappa_{r}+m_r)r_r^2}$,
$\Lambda_9=\frac{{\mu'_r}^2\kappa'_r(1+\kappa'_r)}{(\mu'_r\kappa'_r+m'_r){{r'_{r}}}^2}$, $\kappa$ and $\mu$ denote the fading severity, $m$ represent the shadowing parameters, $r$ denote the rms values of the transmitted signal envelopes of the respective link.
Now, denoting $Z=|\sum^{N_{r}}_{i=1}\alpha_ie^{\phi_{i}}|^2(1-\rho^2)$, the PDF of $Z$ is written as 
$f_{Z}(x)=\frac{1}{\sigma_r}e^{-\frac{x}{\sigma_r}}, x\geq0$, where $\sigma_r=N_r(1-\rho^2)$. Now, using the definition of $\gamma_r$, the CDF of $\gamma_r$ is written as $F_{\gamma_r}(\gamma) 
     =\int_0^\infty\int_0^{(x+\frac{1}{\Bar{\gamma}_r})\frac{\gamma}{\rho^2}}f_Y(y)f_{Z}(x) dx dy.$
 However, for the mathematical simplicity, we approximate $\frac{1}{\bar{\gamma}_r}$ as a Gamma-distributed random variable $Z'$, whose PDF is given by $f_{Z'}(x)=\frac{{a_{r}}^{b_{r}} x^{b_{r}-1}}{\Gamma(b_{r})}e^{-a_{r}x}, x\geq 0$.
Finally, we get
\begin{align}
    \label{pdf4}
       F_{\gamma_r}(\gamma)
       =\int_0^\infty\int_0^{\frac{x\gamma}{\rho^2}}f_Y(y)f_{Z'}(x) dx dy.
\end{align}
Now, utilizing the identities of \cite[Eqs. (2.24.2.2)]{ prudnikov1986integrals} and \cite[Eqs. (7.813.1)]{ edition2007table} into \eqref{pdf4}, the CDF of $\gamma_r$ is derived. 

\bibliographystyle{IEEEtran}
\bibliography{IEEEabrv,asmbBiblio}

\end{document}